\documentclass[aps,prl,superscriptaddress,reprint,onecolumn,notitlepage]{revtex4-1}
\usepackage{xspace}
\usepackage{graphicx}
\usepackage{color}
\usepackage{amsmath, amsthm, amssymb}
\usepackage{natbib}

\newcommand{\ucsd}{Department of Physics, University of California,
                          San Diego, La Jolla CA 92093}
\newcommand{\ricebioen}{Department of Bioengineering, Rice University, Houston, TX}
\newcommand{\ricectbp}{Center for Theoretical Biological Physics, Rice University, Houston, TX}

\newcommand{\ri}{\ensuremath{\vb{r}}^i\xspace}

\newcommand{\qi}{\ensuremath{\vb{q}}^i\xspace}
\newcommand{\betai}{\ensuremath{\beta^i\xspace}}
\newcommand{\betab}{\ensuremath{\bar{\beta}\xspace}}

\newcommand{\rj}{\ensuremath{\vb{r}}^j\xspace}
\newcommand{\rij}{\ensuremath{\hat{\vb{r}}^{ij}}\xspace}
\newcommand{\dij}{\ensuremath{d^{ij}}\xspace}
\newcommand{\poi}{\ensuremath{\vb{p}}^i\xspace}
\newcommand{\eq}{Eq.~}

\newcommand{\fig}{Fig.~}
\newcommand{\dcil}{\ensuremath{D_{\textrm{CIL}}}\xspace}

\newcommand{\vb}[1]{ {\mathbf #1}}

\newcommand{\rb}{\ensuremath{\vb{r}}\xspace}

\newcommand{\meanc}[1]{\ensuremath{\langle{#1}\rangle_c}\xspace}

\newcommand{\ci}{\ensuremath{\textrm{CI}}\xspace}

\begin{document}
\title{Collective signal processing in cluster chemotaxis: roles of adaptation, amplification, and co-attraction in collective guidance}
\author{Brian~A.~Camley}
\affiliation{\ucsd}
\author{Juliane~Zimmermann}
\affiliation{\ricectbp}
\author{Herbert~Levine}
\affiliation{\ricectbp}\affiliation{\ricebioen}
\author{Wouter-Jan~Rappel}
\affiliation{\ucsd}
\begin{abstract}
Single eukaryotic cells commonly sense and follow chemical gradients, performing chemotaxis. Recent experiments and theories, however, show that even when single cells do not chemotax, clusters of cells may, if their interactions are regulated by the chemoattractant.  We study this general mechanism of ``collective guidance" computationally with models that integrate stochastic dynamics for individual cells with biochemical reactions within the cells, and diffusion of chemical signals between the cells.  We show that if clusters of cells use the well-known local excitation, global inhibition (LEGI) mechanism to sense chemoattractant gradients, the speed of the cell cluster becomes non-monotonic in the cluster's size -- clusters either larger or smaller than an optimal size will have lower speed.  We argue that the cell cluster speed is a crucial readout of how the cluster processes chemotactic signal; both amplification and adaptation will alter the behavior of cluster speed as a function of size.  We also show that, contrary to the assumptions of earlier theories, collective guidance does not require persistent cell-cell contacts and strong short range adhesion to function.  If cell-cell adhesion is absent, and the cluster cohesion is instead provided by a co-attraction mechanism, e.g. chemotaxis toward a secreted molecule, collective guidance may still function.  However, new behaviors, such as cluster rotation, may also appear in this case.  Together, the combination of co-attraction and adaptation allows for collective guidance that is robust to varying chemoattractant concentrations while not requiring strong cell-cell adhesion.  
\end{abstract}

\maketitle

\section*{Introduction}

Many individual cells, including white blood cells and bacteria, chemotax -- 
sensing and following gradients of signals.  Some cells, though, aren't 
loners -- they migrate collectively -- and cells traveling in clusters and 
sheets during development must chemotax together.  
Recent experiments \cite
{malet2015collective,theveneau2010collective,bianco2007two} show that 
clusters can have abilities that single cells lack: clusters of those cells 
can follow a gradient even when single cells don't, or move in the opposite 
direction.  One possible explanation for this is the qualitative model of 
collective guidance \cite{rorth2007collective}, in which a cluster of cells 
can gain a direction even though each of its individual cells senses only the 
level of signal, and not its gradient. We have recently introduced a model of 
collective guidance in the context of neural crest cells where the cluster's directionality comes from a regulation of 
contact inhibition of locomotion (CIL) \cite{camley2015emergent}; a related model was introduced for lymphocytes \cite
{malet2015collective}.  However, the field's current understanding of collective guidance and how collective chemotaxis occurs without
single-cell gradient sensing does not account for many aspects of a potential cluster response to a chemoattractant signal.  A minimal model of
 collective guidance is that each cell only reacts to the local chemoattractant and the presence of its neighbors \cite{camley2015emergent}.
More complicated signal processing could take place on the cluster scale if cells communicate with each other to collectively
process the information contained in the chemoattractant gradient.  What experimental signatures would tell us if this were happening,
and how would this signal processing change the efficiency of the cluster's movement?  Can collective signal processing overcome shallow gradients seen {\it in vivo} (e.g. \cite{cai2014mechanical}), amplifying differences in cluster behavior between the front and the back?  In minimal models of collective guidance \cite{camley2015emergent,malet2015collective}, the cluster moves by a tug of war, and is likely under tension.  Nevertheless, collective chemotaxis can also occur in the absence of strong adhesion \cite{theveneau2010collective}.  How does this happen?  We will address all of these questions in this paper.  

Collective chemotaxis and guidance are immediately relevant to both developmental and malignant processes {\it in vivo}.  In particular, we note border cell cluster migration in {\it Drosophila} embryogenesis, in which a small cluster of border cells collectively follows gradients of chemoattractants including PVF1 \cite{montell2012group}.  Neural crest migration in Xenopus also occurs collectively, and requires the neural crest chemoattractant Sdf1 \cite{theveneau2010collective}.  These are two well-studied examples where there is also evidence for the collective guidance hypothesis; however, as collective cell motility is ubiquitous in development \cite{friedl2009collective,vedula2013collective}, collective chemotaxis may be crucial whenever chemotaxis drives cell patterning.  In addition, these topics have an important health relevance in collective cancer motility \cite{friedl2012classifying}, as recent experiments suggest that tumor cell clusters are particularly effective metastatic agents \cite{aceto2014circulating}.  However, our initial focus will be on understanding {\it in vitro} experiments in relatively controlled environments \cite{theveneau2010collective,malet2015collective}, and using these results to develop a useful quantitative framework for the study of collective guidance, which can then be applied to understanding these {\it in vivo} situations.

We first study models of biochemical processing of the chemoattractant signal within the cell cluster, assuming strong cell-cell adhesions as in our earlier model \cite{camley2015emergent}.  We treat the possibility of gradient sensing via cell-cell communication, using a mechanism that allows adaptation to a changing signal level: a local excitation, global inhibition (LEGI) scheme \cite{levchenko2002models}.  We also consider the possibility of cluster-level amplification of a sensed gradient.  With both adaptation and a switch-like amplification, we find that clusters of an optimal size are more efficient at chemotaxing than either smaller or larger clusters.  Amplification of the external signal allows clusters to develop a large velocity even in a shallow gradient.  We argue, based on simple scaling principles, that sufficiently large clusters with only short-range adhesion undergoing collective guidance would be expected to either fragment or become increasingly slow.  

We then show that if the cohesion of a cluster is not controlled by local cell-cell adhesion, but rather by chemotaxis toward a secreted signal (or ``co-attraction" \cite{carmona2011complement}), a cluster of cells can undergo collective guidance by regulation of CIL even if cells are not in continuous contact.  We show how co-attraction and regulated CIL interact in order to create robust chemotaxis.  
In the presence of co-attraction, new behaviors, including persistent cluster rotation, may emerge.  We provide an extensive characterization of the transition to rotation, and how rotation can alter the efficiency of gradient-sensing clusters.

\section*{Model}

\begin{figure*}[h]
\includegraphics[width=160mm]{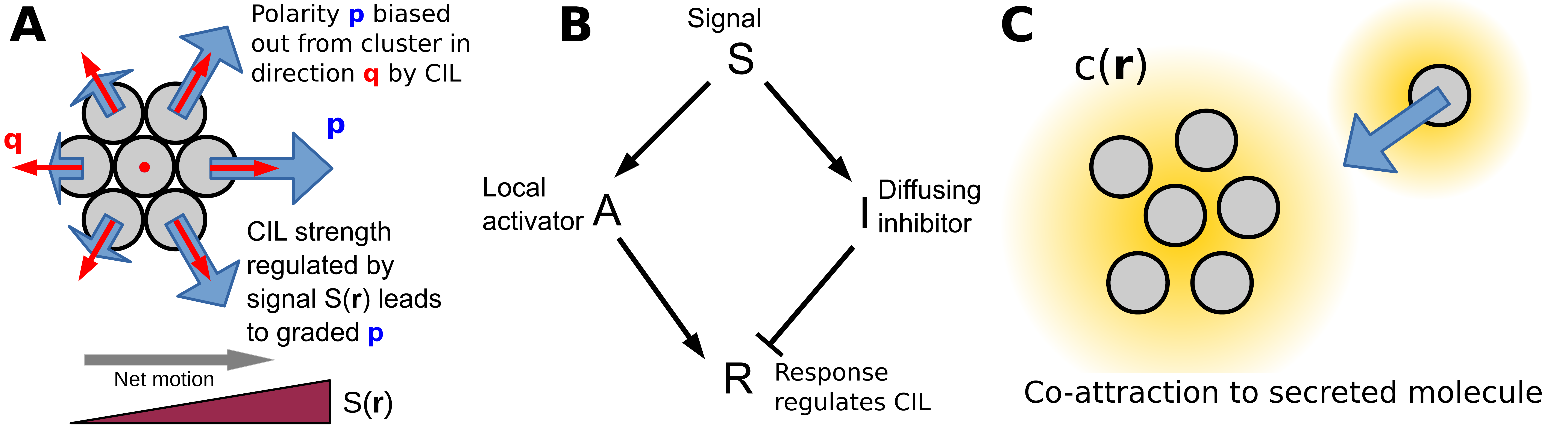}
\caption{\linespread{1.0}\selectfont{}{\bf Summary of model mechanisms}.  (A) Schematic picture of minimal model and origin of directed motion introduced in \cite{camley2015emergent}.  Cell polarities are biased away from the cluster toward the direction $\qi = \sum_{j \sim i} \rij$ by contact inhibition of locomotion (CIL); the strength of this bias is regulated by the local chemoattractant value $S(\rb)$, leading to cells being more polarized at higher $S$.  In our minimal model, we assume the bias strength $\beta$ is directly proportional to $S$, but in models with adaptation and amplification, $\beta$ is controlled by the concentration of the response chemical $R$.  See text for details.  (B) Regulation of CIL strength by signal via a local excitation, global inhibition (LEGI) mechanism. The activator $A$ is localized in each cell, while the inhibitor $I$ may diffuse between contacting cells.  (C) Cluster cohesion may arise from co-attraction, where cells secrete a molecule $c$ which diffuses in the extracellular space.  Individual cells chemotax up the gradient $c(\rb)$. }
\label{fig:schematic}
\end{figure*}

\subsection*{Collective guidance as driven by contact inhibition of locomotion: quick summary and biophysical motivation}

We want to model the collective guidance of a cluster of cells exposed to a chemical gradient $S(\rb)$; we use the experiments of \cite{theveneau2010collective} on neural crest cells as a guide to motivate the features we include, though we expect our results to be more generally applicable as well.  There are four major elements of a model of this process: 1) single-cell dynamics, 2) physical interactions between cells and contact-range effects like contact inhibition of locomotion, 3) the response of the cells to the chemical $S(\rb)$, and 4) chemical communication and signaling between cells.

\textbf{Single-cell dynamics.} We model single cell dynamics using a stochastic particle model that, in the absence of other cells, creates an unbiased persistent random walk even in the presence of a chemoattractant gradient.  Single isolated cells in our model have a behavior that is completely independent of the chemical signal $S(\rb)$.

\textbf{Physical interactions and short-range contact inhibition of locomotion.} We include three short-range interactions between cells: cell-cell adhesion, exclusion of overlap between cells, and contact inhibition of locomotion (CIL).  CIL is a well-known property of many cell types in which cells polarize away from cell-cell contact \cite{carmona2008contact,mayor2010keeping,camley2014polarity,abercrombie1979contact,desai2013contact}.  We model CIL by assuming that the cell polarity, $\vb{p}$, which describes the cell's orientation and propulsion strength, is biased away from directions in which the cell touches another cell.  The result of this assumption is that, consistent with experiments on neural crest cells \cite{theveneau2010collective}, the outside edge cells of a cluster are polarized outward, while cells in the interior are unpolarized (\fig \ref{fig:schematic}A). 

\textbf{Response of cells to chemoattractant.}  For a collective guidance mechanism to drive cluster motility, the polarization must be graded across the cluster.  As the polarization of the cells at the edge is driven by CIL, we assume that CIL is regulated by the chemoattractant concentration $S(\rb)$.  This is motivated by the result of Ref. \cite{theveneau2010collective}, who observe that protrusions on the outside of neural crest clusters are stabilized by the chemoattractant Sdf1.  (The assumption that there is an interaction between chemoattractants and CIL is also supported by other recent experiments \cite{lin2015interplay}, though we do not explicitly model their results here.)  Our earlier minimal model assumed that CIL strength is proportional to the chemoattractant signal $S(\rb)$ \cite{camley2015emergent}.  In this work, we will also treat explicit models of the biochemical processing of the signal within cells, including a local excitation, global inhibition (LEGI) mechanism as well as potential amplification of the chemoattractant signal.  In the LEGI model, the signal creates both activator and inhibitor molecules within the cell, with the activator localized to each cell and the inhibitor free to diffuse between contacting cells via gap junctions.  The activator positively regulates the susceptibility of CIL, and the inhibitor negatively regulates it (\fig \ref{fig:schematic}B).  All of these mechanisms result in cells at higher values of $S$ having a larger susceptibility to CIL, being more polarized, and the cluster moving up the gradient of $S$ (\fig \ref{fig:schematic}A).  

\textbf{Signaling between cells. }  We model a potential ``co-attraction'' between cells as previously seen in neural crest \cite{carmona2011complement,woods2014directional}.  In this mechanism, single cells both secrete a chemical $c$ into the extracellular space and chemotax toward higher levels of $c$ (\fig \ref{fig:schematic}C).  We note that in our model, isolated cells can chemotax toward the secreted co-attractant $c(\rb)$, but do not sense gradients in the signal $S(\rb)$.  This mechanism can provide cohesion to a cluster even if cell-cell adhesion is completely absent.

\subsection*{Mathematical description of model}

We use a two-dimensional stochastic particle model to describe cells exposed to a chemical gradient $S(\rb)$.  
We describe each cell $i$ with a position $\ri$ and a polarity $\poi$.  The cell polarity indicates its direction and propulsion strength: an isolated cell with polarity $\poi$ will travel with velocity $\poi$.  The cell's motion is overdamped, so physical forces like cell-cell adhesion and exclusion change the cell's velocity -- the velocity of the cell is $\poi$ plus the net force the other cells exert on it, $\sum_{j \neq i} \vb{F}^{ij}$.  We model chemically-induced effects like CIL as altering a cell's biochemical polarity $\poi$.  Our model is then:
\begin{align}
\label{eq:position} \partial_t \ri &= \poi + \sum_{j \neq i} \vb{F}^{ij} \\
\label{eq:polarity} \partial_t \poi &= -\frac{1}{\tau} \poi + \sigma {\boldsymbol\xi}^i(t) + \betai \sum_{j \sim i}\rij \\ 
\nonumber  & \; \; \; \; \; \; \; \; \; \; \; \; \; + \chi \frac{\nabla c(\rb^i)}{|\nabla c|} \Theta(|\nabla c|-g_0)
\end{align}
where $\vb{F}^{ij}$ are intercellular forces, e.g. cell-cell adhesion and volume exclusion, and ${\boldsymbol\xi}^{i}(t)$ are fluctuating, temporally uncorrelated noise terms that are Gaussian with $\langle \xi^i_\mu(t) \xi^j_\nu(t') \rangle = 2 \delta_{\mu\nu} \delta^{ij} \delta(t-t')$, where the Greek indices $\mu,\nu$ run over the dimensions $x,y$.  The first two terms on the right of \eq \ref{eq:polarity} are a standard Ornstein-Uhlenbeck model \cite{selmeczi2005cell,vankampen}: $\poi$ returns to zero with a timescale $\tau$, but is pushed away from zero by the fluctuating noise $\boldsymbol\xi(t)$.  This models a cell that has a motion that is only persistent over a time of $\tau$.  

\textbf{Cell-cell forces} We adapt the cell-cell force from \cite{szabo2006phase}
\begin{equation}
\vb{F}^{ij} = \rij
\left\{
	 \begin{array}{lr}
	 v_r \left( \dij - 1 \right), & \; \; \dij < 1 \\
	 v_a \frac{ \dij-1}{D_0 - 1}, & \; \; 1 \le \dij < D_0 \\
	 0                              & \; \; \dij > D_0
	 \end{array}
	 \right.
\end{equation}
where $\dij = |\vb{r}^i-\vb{r}^j|$.  This force is a repulsive spring below the equilibrium separation, an attractive spring above it, and vanishes above $D_0$.  
We will change $v_a$ and $v_r$ in our simulations to move between clusters that are strongly adherent and those with no short-range adhesion (e.g. $v_a = 0$).

\textbf{Contact inhibition of locomotion} We introduced the third term on the right of \eq \ref{eq:polarity} in Ref. \cite{camley2015emergent} to model contact inhibition of locomotion (CIL): the cell's polarity is biased away from cells near it, toward the vector $\qi = \sum_{j \sim i} \rij$, where $\rij = (\ri - \rj) / |\ri-\rj|$ is the unit vector pointing from cell $j$ to cell $i$ and the sum over $j \sim i$ indicates the sum over the neighbors of $i$ (those cells within a distance of $\dcil = D_0$).  
 For cells along the cluster edge, the direction of the CIL bias ($\qi$) points outward from the cluster, but for interior cells $\qi$ is typically smaller or zero (\fig \ref{fig:schematic}a).  Cells around the edge are strongly polarized away from the cluster, while interior cells have weaker protrusions, as observed by \cite{theveneau2010collective}.

The strength of the CIL bias for cell $i$ (i.e. the susceptibility to CIL) is given by $\beta^i$ in \eq \ref{eq:polarity}. This parameter is regulated by the chemoattractant signal $S(\rb)$, as we discuss below.  

\textbf{Co-attraction between cells} The final term on the right of \eq \ref{eq:polarity} is the co-attraction effect: single cells chemotax toward higher levels of $c$.  Here, $\Theta(x)$ is the Heaviside step function, $\Theta(x) = 0$ for $x < 0$ and $\Theta(x) = 1$ for $x > 0$.  This term biases cells to polarize toward increasing $c$, but assumes the strength of this chemotaxis to $c$ is independent of the gradient strength, once the gradient strength is above the threshold $g_0$.  The saturation of polarization is supported by recent experiments in T cells \cite{yang2015dynamic}; modifying this assumption would change how coherent groups of differing numbers of cells are.  We set $g_0 = 10^{-5}$ to be very small; its major role is to prevent division by zero when $|\nabla c|$ is small.  The gradient at the position $\rb^i$, $\nabla c(\rb^i)$, is computed under the assumption that secretion, degradation, and diffusion of $c$ are much faster than all other processes in our model \cite{woods2014directional}, and is found to be ({\it Appendix})
\begin{equation}
\nabla c(\rb^i) = - \sum_{j \neq i} K_1(|\vb{r}^i - \vb{r}^j|/\ell) \rij   
\end{equation}
where $K_1(x)$ is a modified Bessel function of the second kind and the degradation length $\ell$ is set by $\ell^2 = D/k_c$.  We choose $\ell$ to be five cell diameters ($100$ $\mu m$, or five in our simulation units), similar to the value used by \cite{woods2014directional}.  

\textbf{Effect and processing of chemoattractant signal} We model the chemical $S(\rb)$ as regulating a cell's susceptibility to CIL, $\betai$.  A minimal assumption would be that $\betai = \betab S(\rb^i)$ \cite{camley2015emergent}.  This represents the result of \cite{theveneau2010collective} that the cluster chemoattractant Sdf1 stabilizes CIL-induced protrusions \cite{theveneau2010collective}.  However, we will also allow for the possibility that $\betai$ is regulated in a more complex way:
\begin{align}
\betai &= \betab S(\rb^i) \; \; \; \textrm{(minimal model)} \\
\betai &= \betab f(R^i) \; \; \; \textrm{(models with adaptation or amplification)} \label{eq:amp}
\end{align}
where here $R^i$ is the concentration of $R$ molecules in cell $i$; $R$ here is the final read-out of a signal processing network.  We will primarily study a simple, adapting model of response to the signal $S(\rb)$, the local excitation, global inhibition (LEGI) model (Fig. \ref{fig:schematic}B).  We generalize LEGI to cell clusters and show that it creates adaptation and gradient sensing.  In this LEGI model, signal $S$ produces chemicals $A$ and $I$ within each cell with rates $k_A$, $k_I$.  $A$ and $I$ break down with rates $k_{-A}$ and $k_{-I}$.  $A$ remains localized within each cell, but $I$ can be transferred between contacting cells with rate $k_D$.  $A$ upregulates and $I$ downregulates the final output, $R$, which we assume controls CIL, $\beta = \betab f(R)$ (\fig \ref{fig:schematic}B).  Our model, which generalizes \cite{levchenko2002models} to clusters, is then:
\begin{align}
\partial_t A^i &= k_A S(\ri) - k_{-A} A^i \label{eq:activator} \\
\partial_t I^i &= k_I S(\ri) - k_{-I} I^i - k_D n^i I^i + k_D \sum_{j \sim i} I^j \label{eq:inhibitor}\\
\partial_t R^i &= k_R A^i (1-R^i) - k_{-R} I^i R^i  \label{eq:response}
\end{align}
where $n^i$ is the number of neighbors to the $i^{\textrm{th}}$ cell.  \eq \ref{eq:inhibitor} is a reaction-diffusion model on the network of cells \cite{nakao2010turing,othmer1971instability}.  We note that another group has, in a preprint, recently studied a similar LEGI model on cell clusters \cite{mugler2015limits,ellison2015cell}, though in a limited geometry, and without cell-cell rearrangements.  

In Eqs. \ref{eq:activator}-\ref{eq:response}, we have assumed that the inhibitor $I$ is transferred diffusively between cells with a rate $k_D$.  Assuming diffusive transfer between contacting cells is appropriate if $I$ is transferred from one cell's cytosol to the other, e.g. by gap junctions.  Gap junctions modulate neural crest cell motility {\it in vivo} \cite{xu2001modulation,huang1998gap}, making this plausible, though no diffusing inhibitor has yet been identified.  If gap junctions do not form quickly enough, it may be possible to create adaptation by extracellular secretions, similar to the processes involved in quorum sensing in bacteria \cite{waters2005quorum} or via ``transcytosis'' \cite{bollenbach2005robust}.  We also note that we have, in \eq \ref{eq:activator}-\ref{eq:inhibitor}, assumed that the generation of $A$ and $I$ is directly proportional to $S$; this assumes that there is no saturation of the chemosensing receptors on the cell.

Amplification of the chemotactic gradient can be modeled by choosing the function $f(R)$ in \eq \ref{eq:amp}.  Through this paper, we will study a simplified, switchlike form of amplification, so that the response $f(R)$ is a fixed large value if $R$ is above a threshold value, but near-zero if $R$ is below that threshold.  The form we use is $f(R) = g(R/R_0)$, with $g(x) = \frac{1}{2} \left[ 1 + \tanh \left\{(x-1)/\xi\right\} \right]$.  $R_0$ here is the steady-state value of the response $R$ in a constant signal.

\subsection*{Parameter setting}

Throughout this paper, we choose units such that a typical equilibrium cell-cell separation (20 $\mu$m for neural crest \cite{theveneau2010collective}) is unity, and the relaxation time $\tau = 1$ ($\tau$ can be estimated to be 20 minutes in neural crest \cite{theveneau2010collective}).  Within these units, neural crest cell velocities are on the order of $1$, so we choose $\sigma = 1$.  
When we include adaptation, we assume that the kinetics of \eq \ref{eq:activator} and \eq \ref{eq:response} are fast compared with the dynamics of interest, and set them to their steady states, assuming $k_{-R} \gg k_R$ and thus $R^i = A^{i,ss}/I^i(t)$.  We set the diffusion rate $k_D = 4$ in our units, corresponding to a time for equilibration across gap junctions of a few minutes \cite{wade1986fluorescence}.  We set the rates of generation and decay of the inhibitor to be $k_{I} = k_{-I} = 1$; this is discussed more in the adaptation section.  A complete list of parameters is included in the {\it Appendix}, Table A1.

\subsection*{Numerical methods}
We integrate Eqs. \ref{eq:position}-\ref{eq:polarity} and Eqs. \ref{eq:activator}-\ref{eq:response} explicitly with an Euler-Maruyama integrator \cite{kloeden1992numerical}.  The time step varies: for rigid clusters with high adhesion, we choose $\Delta t = 10^{-4}$, and for co-attraction simulations we choose $\Delta t = 5 \times 10^{-3}$.   

\section*{Results}

\subsection*{Review of minimal model of collective guidance in strongly adherent cell clusters}
In our earlier paper \cite{camley2015emergent}, we studied a minimal version of the model described above, with no co-attraction ($\chi = 0$) and no adaptation or amplification, i.e. $\beta^i = \betab S(\rb^i)$.  
We briefly summarize a few results from that paper here, as in some limits, our more complex model will reduce to this model.  Under assumptions of cluster rigidity and slow reorientation, the mean drift of a cluster of cells obeying Eqs. \ref{eq:position}-\ref{eq:polarity} is given by
\begin{equation}
\langle \vb{V} \rangle_c \approx \betab \tau \mathcal{M}\cdot \nabla S \label{eq:shallow}
\end{equation}
with the approximation being true only for shallow gradients, $S(\rb) \approx S_0 + \rb\cdot \nabla S$.  $\meanc{\cdots}$ is an average over the fluctuating $\poi$ but with fixed configuration and orientation of cells $\ri$.  The matrix $\mathcal{M}$ depends only on the configuration of cells; formulas for many cluster shapes and sizes are given in \cite{camley2015emergent}.  We found in particular that the mean cluster velocity $\langle V_x \rangle$ saturates at large $N$.  This arises because we have the difference in signal between the front and the back growing as the cluster radius ($\sim \sqrt{N}$), while the perimeter of the cluster also grows as $\sqrt{N}$.  The force on the cluster then grows as $N$ at large $N$, while the effective friction of the cluster grows independently with the number of cells, as $N$ -- hence the net velocity should behave as $\sim N^{1/2} \times N^{1/2} / N \sim 1$ at large $N$.  We will show later that this assumption will change significantly as we move beyond the minimal model.

We also provided analytic results for the chemotactic index \ci -- a measure of the directionality of the cluster motion.  This is commonly defined as the ratio of the distance traveled in the direction of the gradient (the $x$ direction) to the total distance traveled.  To clarify how we average over many realizations of a path, we define 
\begin{equation}
\ci = \langle V_x \rangle / \langle | \vb{V} | \rangle \label{eq:ci}.
\end{equation} 
For the simple model of \cite{camley2015emergent}, $\ci$ is only a function of $\langle V_x \rangle / \langle \delta V_x^2 \rangle^{1/2}$, where $\langle \delta V_x^2 \rangle = \langle (V_x - \langle V_x \rangle )^2 \rangle = \sigma^2 \tau / N = \langle (V_y - \langle V_y \rangle )^2 \rangle$, independent of cluster geometry.  

\subsection*{Adaptation and amplification in strongly adherent cell clusters}

\textbf{Motivation for adaptation}
With our minimal model of \cite{camley2015emergent}, the cluster velocity is directly proportional to the signal gradient $\nabla S$ (e.g. \eq \ref{eq:shallow}).  This suggests that clusters will slow, and become less directional, if the gradient strength is small -- even if the percentage change in signal across the cell is large.  However, we know that the statistical limits of a single cell's response to a gradient are controlled by the percentage change across the cell \cite{fuller2010external,hu2010physical,hu2011geometry} -- suggesting that the minimal model could be far from optimal.  The minimal model's responses could potentially be improved by internal signal processing to adapt to and amplify the detected chemical signal $S(\rb)$.

\begin{figure*}[htb,floatfix]
\centering
\includegraphics[width=180mm]{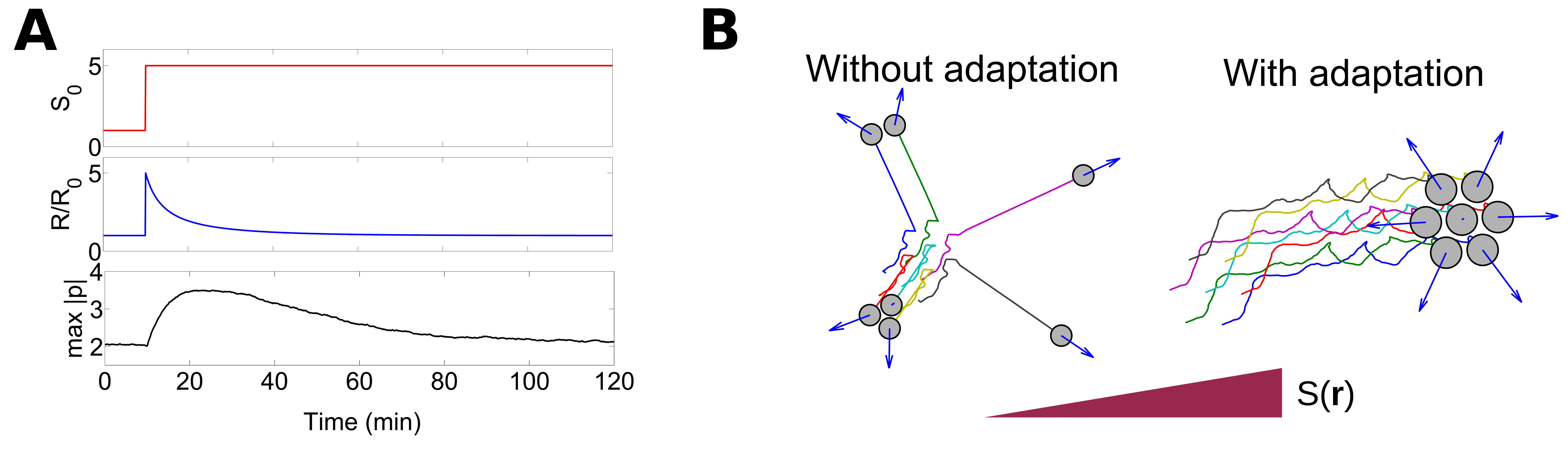}
\caption{\linespread{1.0}\selectfont{}{\bf Signal processing via LEGI leads to perfect adaptation and prevents cluster breakup.}  (A) Simulation of adaption to an increase in a homogenous signal $S_0$.  Both $R$ (which is the same for all cells) and the maximum $|\poi|$ respond and then adapt.  $R_0 = \frac{k_R}{k_{-R}}\frac{k_A}{k_{-A}} \frac{k_{-I}}{k_I}$ is the scale of $R$ and we assume $\beta^i = \betab R^i/R_0$. (B) Weakly adherent clusters without adaptation scatter as they move up the gradient, but those that adapt do not; $v_a = v_r = 23$ in this simulation.  Arrows are polarity $\poi$. }
\label{fig:adapt1}
\end{figure*}

Many cellular responses undergo adaptation, where the response to a signal returns to a baseline level when exposed to a persistently elevated signal \cite{yi2000robust,levchenko2002models,takeda2012incoherent}; adaptation can also allow for easier amplification of a shallow signal.  However, in our minimal model \cite{camley2015emergent}, cell clusters do not adapt to a changing level of signal -- a larger signal leads to a larger effect of CIL, and a larger tension on a cluster.  This suggests that eventually, as strongly adherent cell clusters travel up a chemotactic gradient, they would break apart (\fig \ref{fig:adapt1}B).  This may be reasonable in studying neural crest cells, in which aggregates do separate \cite{theveneau2010collective}, but for strongly adherent clusters like the border cell cluster \cite{montell2012group}, regulating this may be desirable.  

How can a cluster of cells adapt its responses while maintaining a graded response across the cluster?  One answer comes from gradient sensing in single eukaryotic cells: a local excitation, global inhibition (LEGI) model \cite{levchenko2002models,levine2013physics,skoge2014cellular,xiong2010cells}.  In this section, we will study the basic behavior of a strongly adherent cell cluster responding to a chemoattractant signal by LEGI-mediated collective guidance.  Here, we will neglect co-attraction ($\chi = 0$) and set our physical cell-cell interactions to be strong enough that the cluster is highly rigid ($v_r = v_a = 500$).  

\textbf{LEGI on a cell cluster creates perfect adaptation to uniform signals}
The LEGI model of Eqs. \ref{eq:activator}-\ref{eq:response} perfectly adapts to changing uniform signals, as is shown in the analysis of \cite{levchenko2002models} for LEGI in a single cell.  We show this explicitly by finding the steady states of the reaction-diffusion equations, $A^{i,ss} = \frac{k_A}{k_{-A}} S^i$ and
$R^{i,ss} = \frac{A^{i}/I^{i}}{A^{i}/I^{i} + k_{-R}/k_R} \approx \frac{k_R}{k_{-R}} \frac{A^i}{I^i}$, where the approximation holds if the decay rate of $R$ is much faster than its creation rate, $k_{-R} \gg k_R$.  For a signal that is constant in space $S(x) = S_0$,  $I^{i,ss} = \frac{k_{I}}{k_{-I}} S_0$, and thus the steady state $R^{i,ss}(t)$ is independent of $S_0$.  If $S_0$ changes over time, $R$ first increases and then adapts to its steady-state value, as do the cell polarities (\fig \ref{fig:adapt1}A).

\textbf{Ideally adapting clusters develop a velocity that saturates at large cluster size}
The LEGI scheme will create a response in the cells that depends on both $S(\rb)$ and the chemical kinetics of our reaction-diffusion model.  In the limit of fast intercellular diffusion ($k_D \gg k_{-I}$) in a connected cluster, $I^{i,ss} \approx \frac{k_{I}}{k_{-I}}  \overline{S}$, where $\overline{S} = N^{-1} \sum_i S^i$ is the mean signal over the cluster ({\em Appendix}).  In this limit and $k_{-R} \gg k_R$, 
\begin{equation}
R^{i,ss} \approx \frac{k_R}{k_{-R}}\frac{k_A}{k_{-A}}\frac{k_{-I}}{k_I} \frac{S(\ri)}{\overline{S}} \equiv R_0 \frac{S(\ri)}{\overline{S}} \label{eq:response_ss}
\end{equation}
Under these assumptions, $R^i$ develops a profile across the cell proportional to the percentage change in the signal $S(\rb)$ across the cell.  In this limit, if the CIL strength is proportional to $R^i$, $\betai = \betab R^i/R_0$, we find that $\beta^i = \betab S(\ri)/\overline{S}$; the strength of polarization is proportional to the local signal divided by the mean signal over the cluster.  This is highly similar to our minimal assumptions -- the only difference is that the CIL strength $\beta^i$ is rescaled by the mean signal level over the cluster.  Therefore, we would expect \eq \ref{eq:shallow} and its associated results to apply, but with $\nabla S \to \nabla S / \overline{S}$.  In particular, we note that this also predicts that the velocity $\langle V_x \rangle$ will saturate to a characteristic value at large cluster size, as in the minimal model.  However, we caution that in a linear gradient, $\nabla S / \overline{S}$ is not independent of the cluster's position.  For this reason, we generally study adaptation in a shallow exponential gradient, $S(x) = S_0 e^{x S_1}$ so that $\nabla S / \overline{S} \approx S_1$ is constant (as done experimentally in \cite{fuller2010external}).

We note that the LEGI adaptation scheme functions effectively if intercellular diffusion is fast: a cluster can sense a gradient, as in the minimal model, but the response $\beta^i$ no longer grows without limit as the cluster travels up the gradient -- preventing scattering (\fig \ref{fig:adapt1}B).  

\textbf{Slow gap junction kinetics can create a non-monotonic dependence of cluster velocity on cluster size.}  We showed above that in the limit of infinitely fast intercellular diffusion, applying the LEGI adaptation mechanism does nothing more than rescale the response by the mean signal across the cluster.  However, in reality, the approximation of infinitely fast diffusion is not realistic.  
What constraints do gap junction kinetics place on the LEGI mechanism?  Gap junction transfer times are of the order of a few minutes \cite{wade1986fluorescence}; we estimate $k_D \approx 0.2$ min$^{-1}$ ($k_D = 4$ in our units).    
For effective gradient sensing, $I$ must equilibrate over the cluster within the timescale $1/k_{-I}$, i.e. $\alpha \equiv k_{-I}/k_D \ll 1/N$ ({\em Appendix}).   

\begin{figure*}[htb,floatfix]
\centering
\includegraphics[width=180mm]{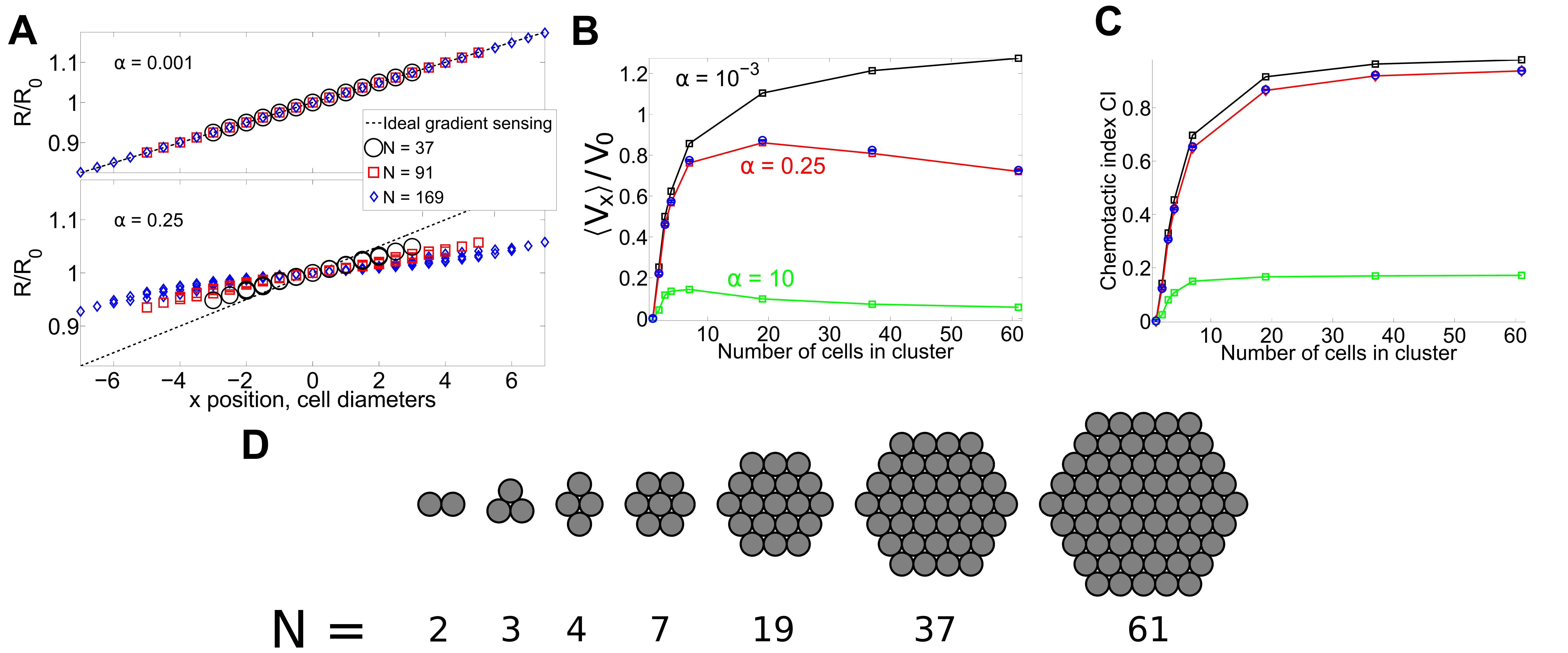}
\caption{\linespread{1.0}\selectfont{}{\bf Gap junction kinetics leads to optimum size for clusters using LEGI adaptation.}   (A) $R^{ss}$ on clusters of different sizes shown for $\alpha = k_{-I}/k_D = 0.25$ (plausible for gap junction transfer), and $\alpha = 10^{-3}$ (near-ideal) in response to a linear gradient.  Dashed line shows the ideal result $R^{ss} = R_0 S(x)/\overline{S}$. (B) Imperfect LEGI gradient sensing creates an optimal cluster size at which speed is maximized.  (C) Chemotactic index of the clusters still increases with increasing $N$.  Squares and lines are results assuming rigid clusters and that \eq \ref{eq:activator}-\ref{eq:response} are at their steady state, while blue circles are full simulations including the reaction dynamics.  $V_0 \equiv \betab\tau S_1$ is the scale of the velocity.  The simulations used to compute (B) and (C) are in an exponential gradient, $S(x) = S_0 e^{S_1 x}$, $S_0 = 1$, $S_1 = 0.025$; $n \ge 2000$ trajectories of $6\tau$ are used for each point. (D) shows the initial cluster shapes assumed in the simulations analyzed in (B) and (C) (though each trajectory starts at a random orientation).    }
\label{fig:adapt2}
\end{figure*}

In principle, $\alpha$ could be decreased arbitrarily and we could reach the fast diffusion limit just by making the inhibitor degradation rate $k_{-I}$ increasingly small.  However, we would not expect $k_{-I}$ to be significantly slower than the cell polarity relaxation rate $\tau^{-1}$.  If $k_{-I} \ll \tau^{-1}$, $R$ will not reach a steady state over the relevant timescales for cell polarity and motility.  We thus expect $k_{-I} \ge \tau^{-1}$.  Based on this estimate and the experimental rates for diffusion via gap junctions, we expect $\alpha \ge 0.25$.  The gap junction kinetics place a strong restriction on LEGI, and clusters have imperfect gradient sensing: $R$ becomes shallower and nonlinear in larger clusters (\fig \ref{fig:adapt2}A).  When this occurs, cluster velocities change.  For $\alpha = 0.25$, mean cluster velocities are non-monotonic in $N$, with a maximum at $N = 19$.  This optimum size can be controlled by changing $\alpha$ (\fig \ref{fig:adapt2}B).  Non-monotonicity can make comparison to experiment difficult -- using cutoffs for ``small'' and ``large'' clusters, as done for some properties in \cite{theveneau2010collective}, could lead to different results depending on the critical cluster size.  Detailed measurements as a function of the cluster radius, as in \cite{malet2015collective}, may be necessary.  

We note that the results in \fig \ref{fig:adapt2}BC show both full simulations of \eq \ref{eq:position}-\ref{eq:polarity} and simplified numerical predictions.  We compute the simplified results by assuming, consistent with the minimal analytic model of \cite{camley2015emergent}, that the cluster is perfectly rigid, and additionally assuming that reaction-diffusion equations are at their steady state.  We can then compute the predicted mean velocity at a particular cluster orientation by $\langle \vb{V} \rangle_c = \frac{1}{N} \sum_i \langle \vb{p}^i \rangle = \frac{1}{N} \sum_{i} \betab f(R^{i,ss}) \vb{q}^i$; the mean velocity $\langle V_x \rangle$ is found by averaging over cluster orientation.  This also defines the chemotactic index of the cluster.  

We show in Fig. \ref{fig:adapt2}C how the chemotactic index \ci changes as a function of the cluster size.  \ci is a measure of the directionality of the cluster, and is defined by $\ci = \langle V_x \rangle / \langle |\vb{V}| \rangle$, where $\vb{V}$ is the cluster velocity (see \eq \ref{eq:ci} and discussion above).  We would expect, based on our results for rigid clusters, that $\ci$ is a monotonic function of $\langle V_x \rangle / \langle \delta V_x^2 \rangle^{1/2}$, where $\langle \delta V_x^2 \rangle = \langle (V_x - \langle V_x \rangle )^2 \rangle = \sigma^2 \tau / N = \langle (V_y - \langle V_y \rangle )^2 \rangle$.  Therefore, if the cluster velocity decreases at large $N$, \ci may still increase if $\langle V_x \rangle$ decreases slower than $N^{-1/2}$.  However, we generally find that \ci reaches a roughly constant value with increasing numbers of cells in the cluster, and this value depends on the parameter $\alpha$ (\fig \ref{fig:adapt2}C).  This suggests that merely the imperfect gradient sensing occurring from the finite gap junction transfer rate will significantly reduce the cluster's chemotactic index, as well as preventing it from increasing significantly as the number of the cells increases.  

We note that a recent preprint has also argued that gap junction limited communication can play a role in limiting chemotactic accuracy in a slightly different context \cite{mugler2015limits}.  

\textbf{Signal amplification can create non-monotonic dependence of cluster velocity on cluster size}

Within chemotaxing single cells, small differences in signal are amplified to large differences in behavior between the cell front and back, allowing efficient migration even in shallow chemotactic gradients \cite{levchenko2002models,levine2013physics,levine2006directional}.  Amplification can also increase cluster motility.  Clusters move via a tug-of-war mechanism -- cells at the cluster back oppose the net motion of the cluster (\fig \ref{fig:schematic}).  If these back cells are suppressed, or the polarization of the cells at the front is amplified, cluster velocity increases. 
 
\begin{figure*}[htb,floatfix]
\centering
\includegraphics[width=180mm]{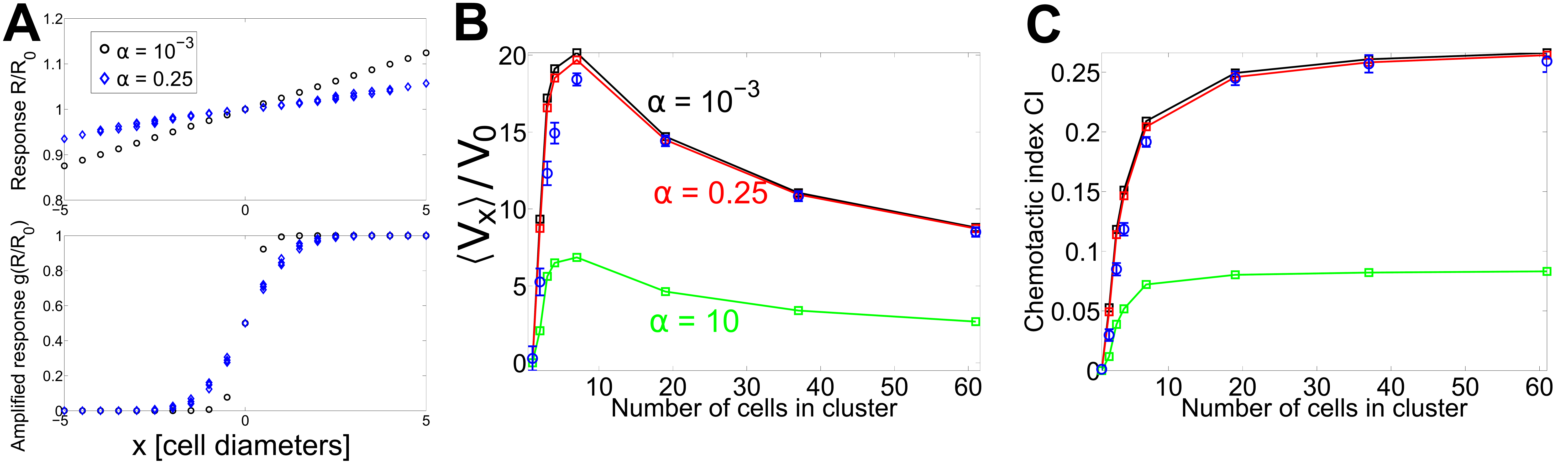}
\caption{\linespread{1.0}\selectfont{}{\bf Amplification can reduce effect of gap junction kinetics, but also creates an optimal size}   (A) $R^{ss}$ and $g(R^{ss})$ on clusters of different sizes.  The use of the switchlike amplification reduces the importance of imperfect gradient sensing, ensuring that the cluster response at the front and the back is the same for either near-ideal ($\alpha = 10^{-3}$) and realistic ($\alpha = 0.25$) gradient sensing. (B) Combination of LEGI gradient sensing and amplification significantly increases cluster speed beyond the typical scale $V_0 \equiv \betab\tau S_1$ and reduces the dependence of cluster speed on $\alpha$, but creates an optimal cluster size at which speed is maximized.  (C) Chemotactic index of the clusters still increases with increasing $N$.  Squares and lines are results assuming rigid clusters and that \eq \ref{eq:activator}-\ref{eq:response} are at their steady state, while blue circles are full simulations including the reaction dynamics.  $V_0 \equiv \betab\tau S_1 $ is the scale of the velocity.  The simulations used to compute (B) and (C) are in an exponential gradient, $S(x) = S_0 e^{S_1 x}$, $S_0 = 1$, $S_1 = 0.025$; $n \ge 2000$ trajectories of $6\tau$ are used for each point.  $\betab = 0.2$ in simulations in this figure -- larger values of $\betab$ can lead to larger deviations from steady-state results. $\betai = \betab g(R/R_0)$, with $g(x) = \frac{1}{2} \left[ 1 + \tanh \left\{(x-1)/\xi\right\}\right]$ and $\xi = 10^{-2}$.}
\label{fig:adapt3}
\end{figure*}

We treat an illustrative but extreme example of amplification in which a cell's response is switchlike, with front cells strongly polarized and back cells suppressed, $\betai = \betab g(R/R_0)$, with $g(x) = \frac{1}{2} \left[ 1 + \tanh \left\{(x-1)/\xi\right\} \right]$.  For $\xi \ll 1$, $\betai \approx \betab$ where $R^i > R_0$ (cluster front), and $\betai \approx 0$ if $R^i < R_0$ (cluster back).
This switchlike response means that the precise value of $R$ is not as crucial as whether it is larger or smaller than $R_0$ (\fig \ref{fig:adapt3}A).  For this reason, 
with strong amplification ($\xi = 10^{-2}$), cluster velocity is, assuming a steady state of $R^i$, much less sensitive to the intercluster diffusion rate $k_D$ (\fig \ref{fig:adapt3}B).  However, the assumption that $R^i$ is at its steady state is not necessarily reasonable for amplified clusters; fluctuating $R^i$ coupled with the nonlinear dependence of $\beta^i$ on $R^i$ above can lead to deviations from the steady-state result (\fig \ref{fig:adapt3}B, blue circles).  In fact, to see reasonable agreement between the steady-state and full kinetics simulations, we had to reduce $\betab$ from its value of 20 in \fig \ref{fig:adapt2} to $\betab = 0.2$ in \fig \ref{fig:adapt3}.

With amplification, cluster velocity increases beyond its usual scale of $V_0 = \betab\tau S_1 $, as the cluster is no longer engaged in a tug-of-war.  However, cluster velocity is still nonmonotonic in cluster size (\fig \ref{fig:adapt3}B), decreasing as $N^{-1/2}$ at large $N$.  This is reasonable, based on a simple scaling argument.  As in the minimal model, polarity is only large along the cluster edge ($\sim N^{1/2}$ cells).  However, with amplification, the polarity strength at the edge is {\it independent of $N$} -- so the mean force driving the cluster scales as $N^{1/2}$, while the friction of the whole cluster scales as $N$, leading to $\langle V_x \rangle \sim N^{-1/2}$.
We see again that the chemotactic index does not significantly increase with cluster size as $N$ increases from 7 to 61 cells (\fig \ref{fig:adapt3}C).  This is expected when $\langle V_x \rangle \sim N^{-1/2}$.  
Other possibilities for amplification (e.g. $\betai = \betab (R^i/R_0)^2$ \cite{levchenko2002models}) will lead to different behaviors for $\langle V_x \rangle$ as a function of $N$.  

\subsection*{Clusters bound only by co-attraction}

Until this point, we have only looked at highly adherent, effectively rigid clusters, where analytic or semi-analytic results are possible.  However, collective cell migration can also occur with a high degree of fluidity and with many cell-cell rearrangements \cite{sepulveda2013collective,angelini2010cell,angelini2011glass,poujade2007collective,petitjean2010velocity,szabo2010collective,szabo2012invasion,vedula2013collective,marel2014flow}.   In addition, we have until now assumed that the only attraction between cells is short-range, representing cell-cell adhesion.  However, neural crest cells also attract one another through chemical secretions, which can control the extent of cluster directionality and cohesion \cite{carmona2011complement,woods2014directional} -- and many other cell types also chemotax toward secretions \cite{mccann2010cell,kessin2001dictyostelium}.  We extend our model to allow for this possibility, and show that clusters of cells that cohere via co-attraction can also be directed by collective guidance.  These clusters need not be rigid, and can have significant re-arrangement or even only transient contacts.  

In this section, we will treat clusters with co-attraction ($\chi \neq 0$), but assume only the minimal model of signal processing, with the CIL susceptibility $\beta^i = \betab S(\rb^i)$.  

\textbf{Clusters with only transient cell-cell contacts can still chemotax effectively}

\begin{figure*}[htb,floatfix]
\centering
\includegraphics[width=180mm]{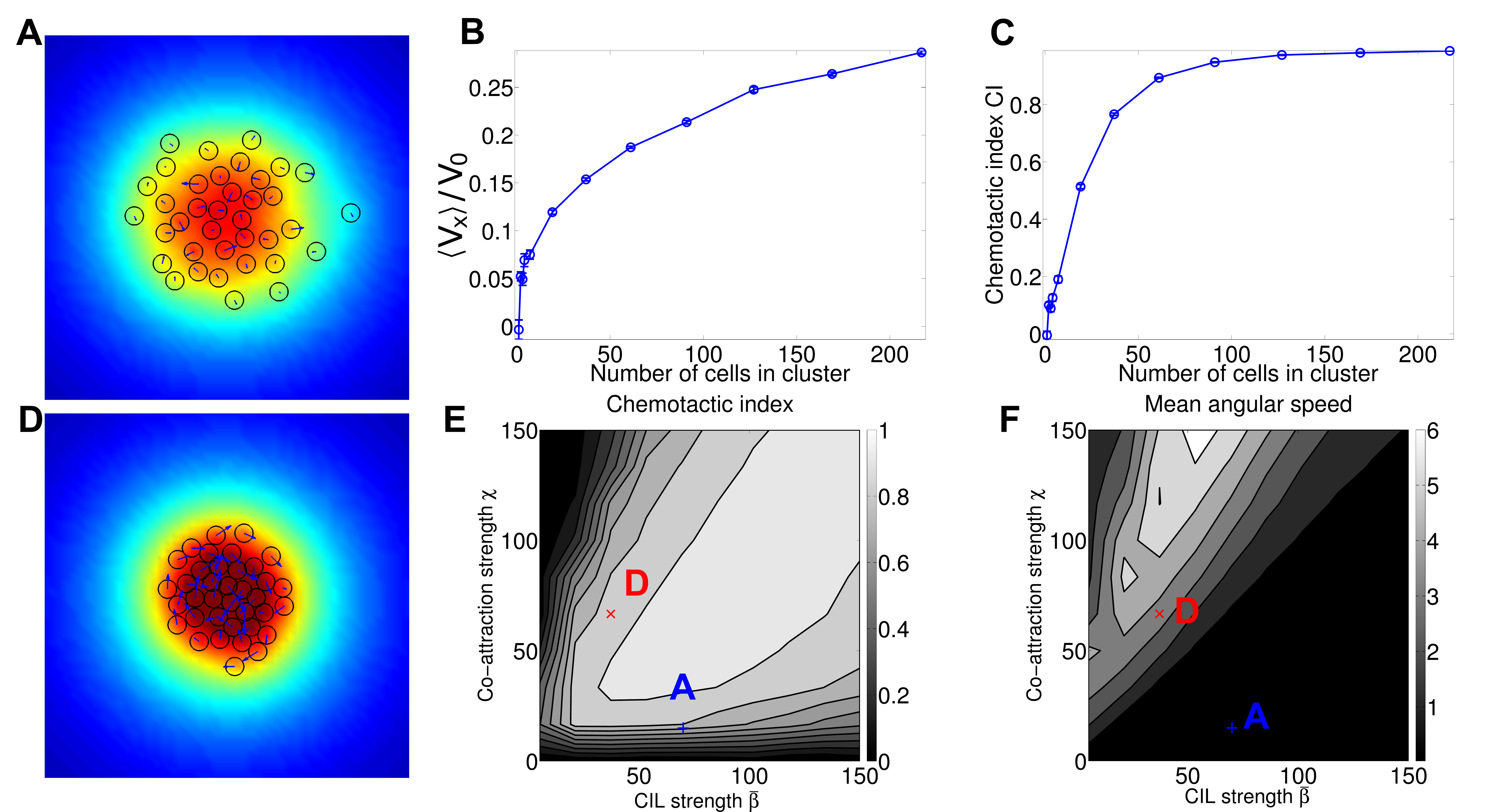}
\caption{\linespread{1.0}\selectfont{}{\bf Co-attraction and graded CIL can create directed motion.}  (A) is a representative snapshot of a chemotaxing cluster loosely bound by co-attraction, with $\betab = 70$ and $\chi = 15$.  (B) and (C) show the velocity and chemotactic index of a cluster with these parameters as a function of number of cells in the cluster.  Both cluster velocity and CI increase with increasing cluster size;  $V_0 = \betab\tau |\nabla S|$.  (D) is a snapshot of a rotating chemotaxing cluster with stronger co-attraction and weaker CIL ($\betab = 37.2$ and $\chi = 83.3$).  In (A) and (D), the color map is the co-attractant field $c(\rb)$, the blue arrows are the cell polarity $\poi$, and the cells are drawn as black circles.  (E) Phase diagram of chemotactic index of clusters of $N = 37$ cells.  CI increases when both co-attraction $\chi$ and CIL strength $\bar{\beta}$ are increased.  (F) Phase diagram of mean angular speed $\langle | \Omega | \rangle$ of clusters of $N = 37$ cells.  Clusters with sufficiently high co-attraction develop rotational motion.  Points corresponding to the simulations shown in (A) and (D) are marked on the phase diagrams of (E) and (F).  
Throughout this figure, the degradation length $\ell = 5$ cell diameters.  $n = 100$ trajectories of length 50$\tau$ are used for each point of the phase diagrams in (E) and (F), which are contour plots based on a $10\times10$ sample of the space $\betab \in [5,150], \chi \in [0,150]$.  Varying numbers of trajectories are used for each point in (B) and (C) ranging from $n = 200$ for $N \le 91$ to $n = 10$ for $N = 217$.  $\Delta t = 0.005$, $v_a = 0, v_r = 100$, and $|\nabla S| = 0.025$ throughout this figure. }
\label{fig:coa}
\end{figure*}

We have, in the sections above and in \cite{camley2015emergent}, studied one extreme limit of cell cluster cohesion: highly cohesive clusters linked by short-range adhesion.  We now study the opposite limit: no short-range adhesion of any sort ($v_a = 0$), with the cohesion of the cluster being provided solely by chemotaxis to the secreted molecule $c$.  Cluster chemotaxis via collective guidance can still function even in this limit (\fig \ref{fig:coa}A-C, Movie S1).  

\textbf{Loosely bound cluster chemotactic velocity increases with cluster size}
Some qualitative results of our minimal, rigid model are recapitulated in the model with co-attraction, though there are important differences.  Larger clusters are generally faster and more efficient (\fig \ref{fig:coa}B,C).  Mean velocity increases sublinearly with cluster size.  However, unlike our studies of rigid clusters, we do not observe the speed of the cluster saturating with cluster size.  Why? The saturation of the cluster velocity for a rigid cluster in the minimal model \cite{camley2015emergent} (or with ideal adaptation, as above) arises because we balance a force due to CIL that is exerted on the edge of the cluster ($\sim \sqrt{N}$ cells), and increases linearly with the radius of the cluster ($\sim \sqrt{N}$), hence increasing as $N$, with a drag that comes from a linear combination of all the individual cells ($\sim N$).  In the mechanism with co-attraction, CIL acts at any cell-cell collision -- and these are not limited to the edge of the cluster (\fig \ref{fig:coa}A).

We note that the co-attraction simulations in \fig \ref{fig:coa} assume the minimal model of $\beta^i = \betab S(\rb^i)$, with no adaptation or amplification (we study the interaction of the LEGI adaptation mechanism and co-attraction later).  For this reason, as the cell cluster travels up the gradient, the mean value of $\beta$ on the cluster increases, which will change the morphology and dynamics of the cluster.  The value of the chemotactic index we present is averaged over the time from $12.5\tau$ to $50\tau$ after the simulation is initialized; changing this averaging range does not qualitatively change the results in \fig \ref{fig:coa}.  

\textbf{Chemotactic index, cluster rotation depend on balance of co-attraction and CIL}
As the degree of co-attraction increases, clusters may also develop a persistent rotational motion while they chemotax (\fig \ref{fig:coa}D, Movie 2).  This is consistent with other simulations that show that self-propelled particles with long-range interactions from chemotaxis or other sources can develop a vortex state \cite{czirok1996formation,levine2000self,d2006self}.  We note, however, that this vortex state arises without an explicit effect aligning cell polarities, and can occur even without CIL.  We provide a simple explanation for the emergence of cluster rotation in the next section.  

We show a phase diagram of the cluster chemotactic index as well as the mean angular speed in the cluster in \fig \ref{fig:coa}E,F.  The chemotactic index is generally maximized when the co-attraction strength $\chi$ and CIL strength $\betab$ are similar, and can be increased by simultaneously increasing $\chi$ and $\bar{\beta}$.  We can understand many of these results intuitively.  Increasing the co-attraction increases the number and duration of interactions, and since the chemotactic response to the signal $S$ emerges from cell-cell interactions, this increases chemotactic efficiency.  Increasing $\betab$ increases the (graded) polarization of the cells due to CIL, and hence the chemotactic index -- but unless $\chi$ also increases, the increase in CIL causes the cluster density to decrease, reducing the number of interactions.  However, increasing $\chi$ to be much larger than $\betab$ leads to both rotation and a decreased chemotactic index.  We emphasize that \fig \ref{fig:coa}C,E plot the {\it cluster} chemotactic index; for loosely bound clusters, especially with rotation, the chemotactic index of an individual cell can be very different from that of the cluster.  

\textbf{Cluster rotation emerges at large co-attraction via a pitchfork bifurcation}

\begin{figure*}[htb,floatfix]
\centering
\includegraphics[width=120mm]{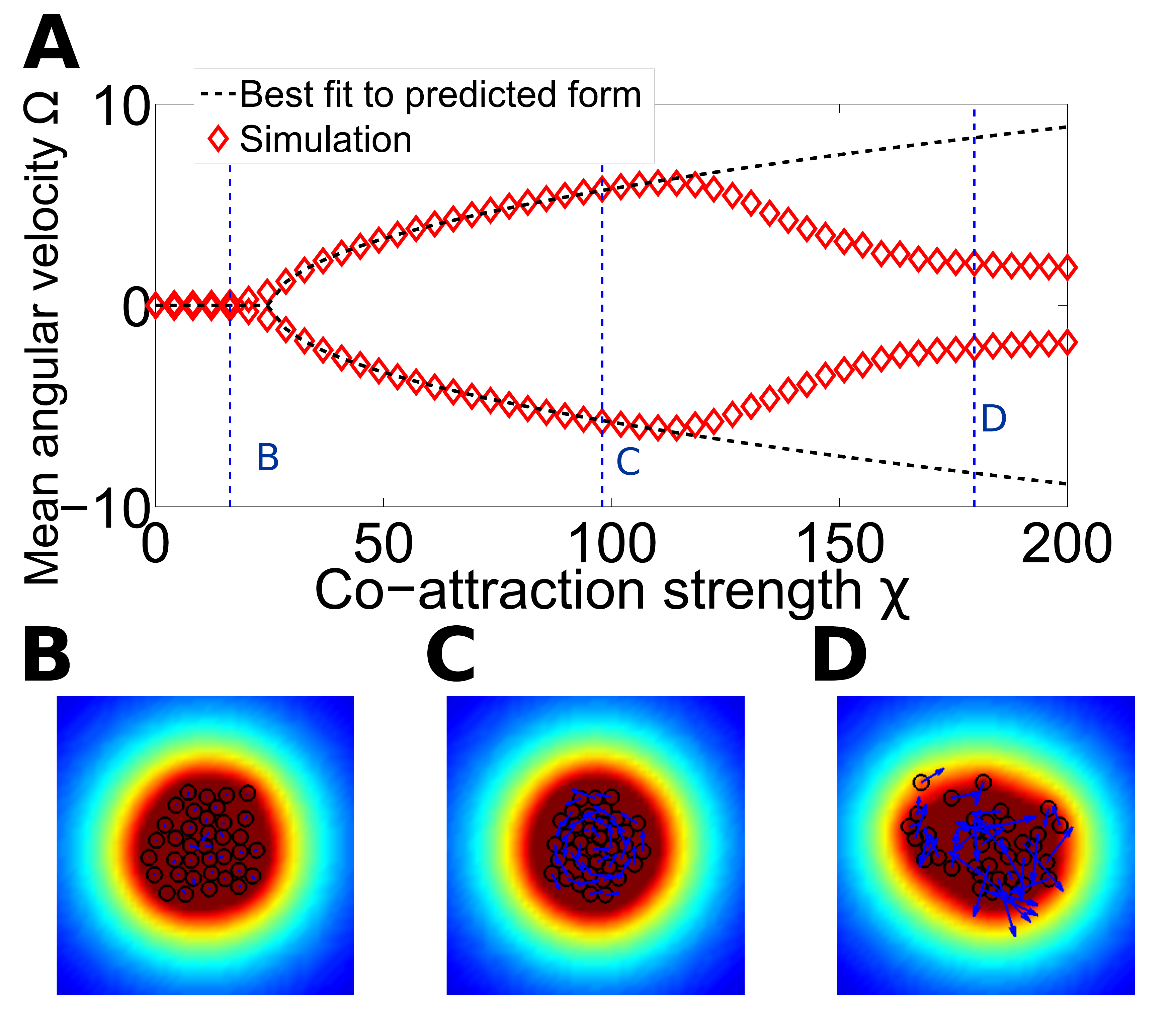}
\caption{\linespread{1.0}\selectfont{}{\bf Cluster rotation with co-attraction emerges as a pitchfork bifurcation}  (A) The mean angular velocity for our full simulation of clusters under co-attraction is in good agreement with the general form of \eq \ref{eq:transition}.  These trajectories are simulated for $N = 37$ cells with $\betab = 35$. The two branches in the simulation (A) are the means over those clusters with positive (negative) angular velocity.  $n = 200$ simulations of length 50$\tau$ are used for each value of $\chi$.  The best fit parameters are $\Omega_0 = 0.671$ and $\chi_c = 25.6$.  $|\nabla S| = 0$ in these simulations.  We highlight three points and show typical simulations in (B), (C), and (D).  (B) $\chi = 16.3$, co-attraction is weak, and contacts are only transient (C), $\chi = 98.0$, and the cluster becomes rigid and rotates (D), $\chi = 179.6$, and co-attraction is so strong that cells squeeze past each other, leading to looser clusters.  In (B)-(D), the color map is the co-attractant field $c(\rb)$, the blue arrows are the cell polarity $\poi$, and the cells are drawn as black circles.  The vector field $\poi$ has a consistent scale in (B)-(D) -- the magnitude of $\poi$ is much larger in (D) due to the higher value of the co-attraction strength $\chi$.}
\label{fig:bifurcation}
\end{figure*}

We can provide a simple argument for how co-attraction can lead to cluster rotation, deriving some results for a simplified version of our model; these results are surprisingly effective in characterizing the full model (\fig \ref{fig:bifurcation}).  The three crucial elements of our simplified model are the finite polarity relaxation time $\tau$, the rigidity of the cluster, and the tendency of co-attraction to polarize cells toward the center of the cluster.  We treat the cluster as rigid even though there is no short-range adhesion, because at sufficiently high co-attraction strength $\chi$, the cluster will be held together by co-attraction, as in \fig \ref{fig:coa}D, and it effectively rotates as a rigid body.  We describe a circular cluster, and for simplicity, assume that co-attraction and contact inhibition only affect the outer layer of cells; this assumption can easily be generalized without significant changes.  We also consider the limit of deterministic motion, where the fluctuating noise in our polarity equations vanishes, $\sigma \to 0$.  We then write
\begin{align}
\partial_t \vb{p}^i &= -\tau^{-1} \vb{p}^i + \Gamma \hat{\vb{n}}^i \\
\partial_t \theta &= \mu_\theta \sum_i \vb{p}^i \cdot \hat{\vb{t}}^i
\end{align}
Here, $\hat{\vb{n}}^i$ is the outward normal to the circular cluster at cell $i$, $\hat{\vb{t}}^i$ is the tangent to the circle, and $\Gamma$ indicates the bias arising from a combination of CIL and co-attraction; $\Gamma$ may be either positive or negative.  $\theta$ is the angle of rotation of the cluster, which is driven by the tangential force exerted on it, $\sum_i \vb{p}^i \cdot \hat{\vb{t}}^i$; $\mu_\theta$ is a mobility.  We can write the normal and tangential directions as $\hat{\vb{n}}^i = \left(\cos \left[\phi_i + \theta\right],\sin\left[\phi_i + \theta\right]\right)$, $\hat{\vb{t}}^i = \left(-\sin \left[\phi_i + \theta\right],\cos\left[\phi_i + \theta\right]\right)$, leading to $\frac{d}{dt} \hat{\vb{n}}^i = \hat{\vb{t}}^i \dot{\theta}$ and $\frac{d}{dt} \hat{\vb{t}}^i = -\hat{\vb{n}}^i \dot{\theta}$.  Here, $\phi_i$ is the angular position of the cell when the cluster is at rest at $\theta = 0$.  Writing $\vb{p}^i = \hat{\vb{n}}^i p_n^i + \hat{\vb{t}}^i p_t^i$, we find
\begin{align}
\dot{p}_n^i - \dot{\theta} p_t^i &= -\tau^{-1} p_n^i + \Gamma  \\
\dot{p}_t^i + \dot{\theta} p_n^i &= -\tau^{-1} p_t^i  \\
\dot{\theta} &= \mu_\theta \sum_i p_t^i
\end{align}
We see immediately two steady states.  The first steady state has no rotation, $\dot{\theta} = 0$, $p_t^i = 0$, and $p_n^i = \Gamma \tau$.  The second steady state has constant rotational motion, $\dot{\theta} = \Omega$, $p_n^i = -\frac{1}{N_p \tau \mu_\theta}$, and $p_t^i = \frac{\Omega}{N_p \mu_\theta}$, where $N_p = \sum_i 1$ is the number of cells on the edge, and $\Omega = \pm \left[-\tau^{-2} - \Gamma N_p \mu_\theta \right]^{1/2}$.  For $\Gamma < -1/\left(\tau^{2} N_p \mu_\theta\right)$, $\Omega$ is real.  Linearizing around the no-rotation steady state, $p_n^i = \Gamma \tau + \delta_n$, $p_t^i = \delta_t$, we find that the non-rotating state becomes unstable for $\Gamma < -1/\left(\tau^{2} N_p \mu_\theta\right)$.  We thus expect that we see rotation once $\Gamma$ becomes sufficiently negative -- i.e. the co-attraction becomes strong.  In our larger model, the effect biasing our polarities toward the center of the cluster arises only from co-attraction -- so we expect that in our full model, the role of $\Gamma$ will be played by $-\chi$.  In this case, we would expect
\begin{equation}
\Omega = \begin{cases} 
      0 & \chi\leq \chi_c \\
      \pm \Omega_0 \sqrt{\chi-\chi_c} & \chi > \chi_c
   \end{cases} \label{eq:transition}
\end{equation}
where $\Omega_0$ and $\chi_c$ will depend on the details of the model, e.g. the extent of CIL and the number of cells.  We show in \fig \ref{fig:bifurcation} that \eq \ref{eq:transition} is a very good description of how our full model with co-attraction transitions to rotation.  Clusters transition to rotation at a critical $\chi_c$, which for the parameters of \fig \ref{fig:bifurcation} is found to be $\chi_c \approx 25.6$, and the angular velocity of clusters increases as $\sqrt{\chi-\chi_c}$ as $\chi$ increases.  However, for $\chi$ large enough, cells become so strongly polarized inward that the cluster breaks apart, and the angular velocity decreases, and \eq \ref{eq:transition} no longer holds.  

We also emphasize that this effect occurs in part because, when the cluster rotates as a rigid body, the polarity $\vb{p}^i$ is constant in the frame of the substrate, rather than the frame of the rotating cluster.  This effect may therefore be modified or absent in models that resolve, e.g. torques on individual cells or full cell structure.

\begin{figure*}[htb,floatfix]
\centering
\includegraphics[width=120mm]{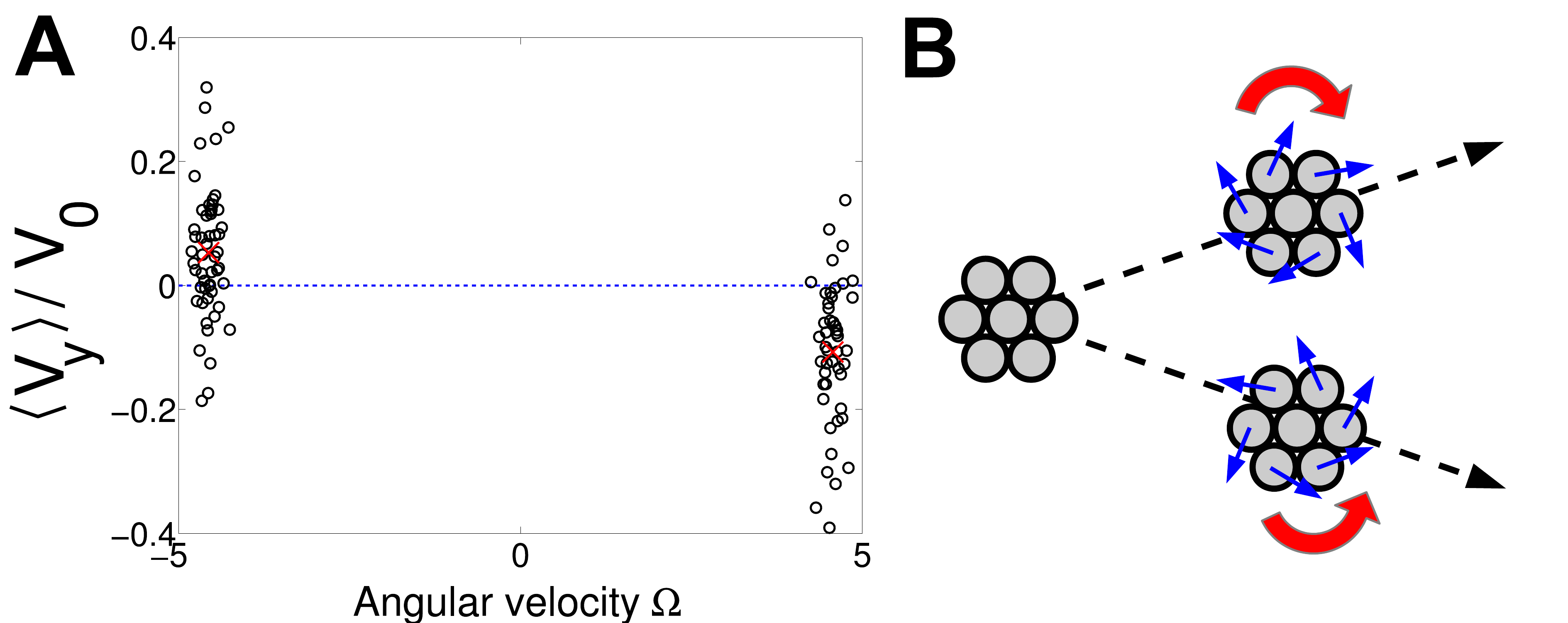}
\caption{\linespread{1.0}\selectfont{}{\bf Rotating clusters develop a handedness-dependent drift perpendicular to the gradient}  (A) The mean drift velocity $\langle V_y \rangle$ for 100 trajectories is plotted as a function of that trajectory's mean angular velocity (black circles).  The mean drift velocities for the clusters with positive and negative angular velocities are plotted as red crosses.  These trajectories are simulated for $N = 37$ cells with the parameters of \fig \ref{fig:coa}D, with the averages evaluated over the time from $12.5\tau$ to $50\tau$ after the simulation is initialized, as in Fig. \ref{fig:coa}D. (B) Schematic picture of cluster trajectories showing that even if the motion is deterministic, the rotation-drift coupling can change the cluster chemotactic index.}
\label{fig:rot}
\end{figure*}

\textbf{Rotating clusters develop chirality-dependent drift} When the cluster develops a spontaneous rotation, this changes the underlying symmetries of our problem.  Ordinarily, for a roughly circular cluster, our system is symmetric with respect to inversions in the $y$ direction (the direction perpendicular to the chemoattractant gradient).  This suggests that the net drift in the $y$ direction should be zero.  However, the rotation of the cluster gives it a handedness that breaks the $y$ inversion symmetry -- allowing a net drift in the $y$ direction, whose sign would then depend on the handedness of the rotation.  We show in \fig \ref{fig:rot}A that this is the case.  This gives a possible explanation for why, in \fig \ref{fig:coa}E,F, the chemotactic index of clusters decreases as cluster rotation begins.  As shown schematically in \fig \ref{fig:rot}B, even if the cluster's motion is completely deterministic, a systematic drift velocity in the $y$ direction will reduce the cluster's chemotactic index, because the mean motion no longer is purely in the $x$ direction.  

What is the origin of this drift?  We cannot explain this result in terms of the assumptions that led to \eq \ref{eq:transition} -- if we extend this model to have a position-dependent $\Gamma$, it will predict that the drift in the y direction is zero.  We suspect that this drift in our full model largely arises from co-attraction.  Small distortions in the cluster arising from the rotation can lead to a bias for cells that would otherwise see relatively weak gradients.  However, we note that this particular mechanism requires cells that are sensitive to weak gradients in the co-attractant, and then amplify this into larger motion.  We have found that handedness-dependent drift in our model is sensitive to details such as our assumption that cells respond equally strongly to weak and strong gradients of the co-attractant $c(\rb)$.  We nevertheless want to highlight the chirality-dependent drift as a feature that is not excluded on symmetry grounds and may arise more robustly in other models.

\textbf{Combination of co-attraction and adaptation is similar to co-attraction in the minimal model}

\begin{figure*}[htb,floatfix]
\centering
\includegraphics[width=180mm]{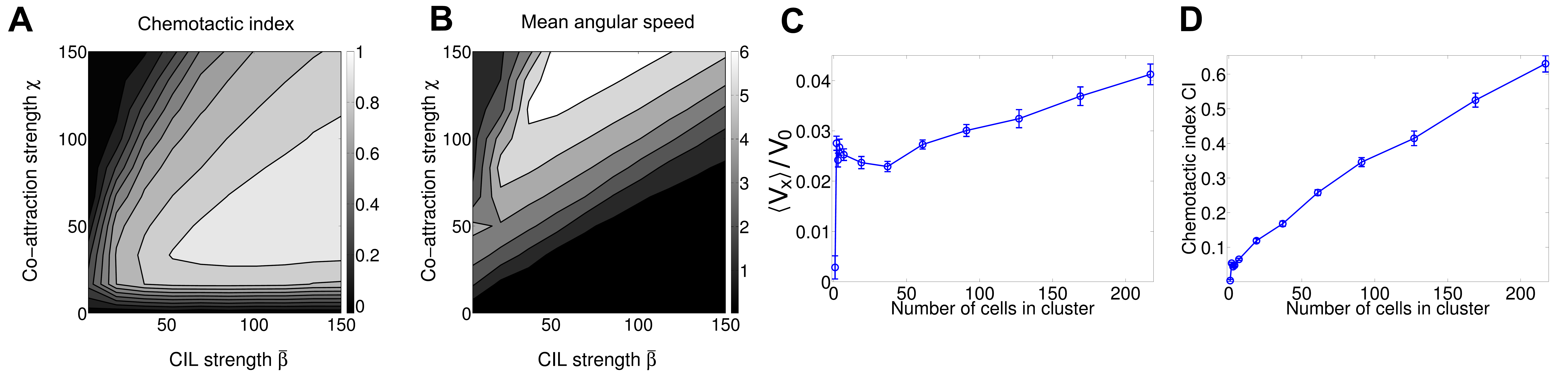}
\caption{\linespread{1.0}\selectfont{}{\bf Combination of co-attraction and adaptation does not qualitatively change phase diagram} (A) Phase diagram of chemotactic index of clusters of $N = 37$ cells.  CI increases when both co-attraction $\chi$ and CIL strength $\bar{\beta}$ are increased, as in \fig \ref{fig:coa}E.  (B) Phase diagram of mean angular speed $\langle | \Omega | \rangle$ of clusters of $N = 37$ cells.  Clusters with sufficiently high co-attraction develop rotational motion, as in \fig \ref{fig:coa}F.  (C) Velocity in gradient direction increases, with a dip, as a function of cluster size, but is significantly smaller than the nominal scale $V_0 \equiv \betab \tau S_1$.  (D) Chemotactic index increases with increasing cluster size.  
Throughout this figure, the degradation length $\ell = 5$ cell diameters.  $n = 100$ trajectories of length 50$\tau$ are used for each point of the phase diagrams in (A) and (B), which are contour plots based on a $10\times10$ sample of the space $\betab \in [5,150], \chi \in [0,150]$.  $\Delta t = 0.005$, $v_a = 0, v_r = 100$. The gradient is exponential, $S(x) = S_0 e^{S_1 x}$, $S_0 = 1$, $S_1 = 0.025$.  (C) and (D) are evaluated at $\betab = 70$, $\chi = 15$.}   
\label{fig:coaadapt}
\end{figure*}

Here, we study our complete model, including both the LEGI adaptation mechanism and co-attraction.  We show the phase diagram for how the chemotactic index \ci and the mean angular speed depend on the strength of CIL $\betab$ and the strength of co-attraction $\chi$ in \fig \ref{fig:coaadapt}AB.  As in our model with co-attraction, but no adaptation, we see rotation at large $\chi$ and a chemotactic index that is maximized at large $\chi$ and $\beta$ (though reduced when rotation occurs).

When we plot the cluster velocity (\fig \ref{fig:coaadapt}C), we note two features.  First, we see that there is an apparent dip in velocity for intermediate numbers of cells, but then an increase at large cell numbers.  This is perhaps not surprising given the combination of adaptation, for which we see a decrease in cluster velocity at large $N$ (\fig \ref{fig:adapt2})B, and co-attraction, where in the absence of adaptation, the cluster velocity increases consistently with increasing $N$ (\fig \ref{fig:coa}B).  Secondly, the cluster speed is significantly smaller than seen in \fig \ref{fig:coa}B, even though the co-attraction and CIL parameters have remained the same.  This is a natural consequence of having cells only in transient contact, as the LEGI mechanism takes time to create a gradient in $R^i$ across an area of cells in contact.  Depending on the time required to form robust connections between cells, this effect could be aggravated -- suggesting adaptation via LEGI may be more difficult for cells with transient junctions.  However, we do see that as the number of cells increases, the cluster is still efficiently directed and \ci increases (\fig \ref{fig:coaadapt}D)

\section*{Discussion}

In our earlier work \cite{camley2015emergent}, we provided a minimal quantitative model that embodied the collective guidance hypothesis \cite{rorth2007collective,theveneau2010collective} and provides a plausible initial model for collective chemotaxis when single cells do not chemotax.  However, this model made two fairly strict assumptions: first, that the cell does not perform any relevant internal processing of the chemoattractant gradient, and second, that the cluster was highly adherent with cohesion from short-range interactions.  In this paper, we have relaxed both of these assumptions.

We find, consistent with our earlier model and the experimental results of \cite{theveneau2010collective}, that small clusters of cells can chemotax, even if single cells cannot.  However, while our minimal model predicts that both velocity and chemotactic index increase as cluster size increases, we find that adaptation to and amplification of the chemoattractant signal can lead to cluster velocity to decrease and chemotactic index to saturate at a value less than one at large cluster sizes.  This effect can arise purely from the finite rate of gap junction mediated transfer of chemicals between contacting cells.  However, even if gap junction transfer is fast, if the cell's response to the signal is switchlike, cluster velocities can be non-monotonic.  This non-monotonic behavior is a sign of the way in which the cluster processes the chemoattractant signal.  

How does this compare to experimental results?  Theveneau et al. \cite{theveneau2010collective} find that chemotactic indices of small (2-3 cell) and large clusters are similar, but do not observe a large variation in cluster speed.  Within our models with adaptation and amplification in adherent clusters, we see that chemotactic indices of 7 cell clusters and 61 cell clusters are often similar -- but we generally observe that 2-3 cell clusters have smaller chemotactic indices.  This is in part because of orientational averaging -- under our assumption that single cells do not chemotax, pairs of cells have a chemotactic behavior that varies strongly with orientation \cite{camley2015emergent}.  However, we highlight some difficulties with a direct comparison between model and experiment in this case.  First, if cells have a distribution of CIL strengths $\betab$, we might expect that cells with higher CIL strength $\betab$ form smaller clusters; thus the strength of the collective guidance mechanism could be different between small and large clusters.  Secondly, the concentration of the chemoattractant Sdf1 is not well-characterized in the bead assay used by \cite{theveneau2010collective}; more extensive studies using microfluidic chambers would aid in pinpointing differences between our model and experiments in neural crest.  Non-monotonicity can also make experimental results difficult to interpret.  Within our model with adaptation and amplification, a given cluster could have either a larger or smaller velocity than a smaller cluster, depending on the sizes of the clusters and the details of the adaptation mechanism (e.g. the rate of diffusion of the inhibitor).  This suggests that solely comparing large and small clusters could potentially be misleading, and in general more detailed experiments as a function of cluster size are needed.  

How generic is the result that sufficiently large tightly bound clusters decrease in speed?  We have found that this occurs both with adaptation and switch-like amplification, but not with the minimal model of \cite{camley2015emergent}.  We expect this behavior to be relatively broadly present in sufficiently large, and sufficiently tightly bound clusters.  In collective guidance in tightly adherent clusters, there is a tug-of-war, and the cluster is under tension, and would tear itself apart in the absence of the strong adhesion.  In order for the cluster to maintain its speed as the number of cells increases, which increases friction between the cluster and the surface, the difference in protrusion strength between the front and back must also increase.  In adaptation and switchlike amplification, this increase in protrusion strength is slow or absent, and the cluster slows at large $N$.  However, if the protrusion strength $\beta^i$ increases with cluster size, it will eventually overcome cell-cell adhesion -- and we expect the cluster to eventually scatter, as in \fig \ref{fig:adapt1}B.  Generically, large tightly adherent clusters will either scatter or slow. 

Our model for strongly adherent clusters has assumed that the primary driver of the dynamics of strongly adherent clusters are cells at the edge.  This is consistent with the observations of \cite{theveneau2010collective} on neural crest, who observe that only the edge cells develop strong protrusions: there are no cryptic protrusions.  In small (2-30 cells) adherent clusters of epithelial cells, traction forces are also mainly found to be large at the edge \cite{maruthamuthu2011cell,mertz2012scaling,mertz2013cadherin}.  However, in other systems, most notably the classic example of migration of large epithelial sheets in a wound healing geometry, traction forces are also exerted significantly away from the edge \cite{trepat2009physical}.  If these interior traction forces are relevant, we would expect the scaling of cluster velocity with cluster size to be significantly altered.  Ultimately, experimental traction force measurements  may be crucial in determining whether our assumption of edge-driven dynamics is appropriate; however, this assumption is consistent with the currently available data.  

Adaptation in single-cell chemoresponse is a ubiquitous and well-tested principle, but its existence is not established for clusters of neural crest or lymphocytes; applying a step response would be a straightforward test of adaptation, and we would expect protrusions and traction forces to peak and then adapt (\fig \ref{fig:adapt1}A).
Our model of adaptation in the cell cluster is a LEGI model that allows cells and cell clusters to adapt to changing levels of chemoattractant and still gradient sense, but requires communication between cells.  We have suggested that this communication may arise from diffusion through gap junctions, and our results suggest that in some cell types, gap-junction mediated gradient sensing across the cluster may be effective -- though with characteristic effects on the cluster velocity, as discussed above.  As noted earlier, our hypothesis of gap junction mediated communication is consistent with the experimental observation that gap junctions modulate neural crest cell motility {\it in vivo} \cite{xu2001modulation,huang1998gap}.  We also note that a recent preprint has independently suggested that gap junctions play a role in gradient sensing and proposed a similar LEGI model \cite{ellison2015cell,mugler2015limits}, though solely in one dimension, and without any effects of CIL, cell motility, or amplification.  

In this paper, we also modeled the possibility that the coherence of the cluster is not provided by strong physical adhesion, but rather by chemoattraction to a secreted signal, i.e. co-attraction.  This co-attraction mechanism is known to be relevant in neural crest \cite{carmona2011complement} (and see also the model \cite{woods2014directional}).  Our model with co-attraction also shows that, consistent with experiments on neural crest \cite{theveneau2010collective}, that the collective guidance mechanism proposed here can guide cells even with only transient contacts.  However, we also see that if co-attraction is too large, new emergent behaviors can appear, including cluster rotation.  Persistent rotation of cell clusters is not observed in neural crest, and only transient rotations appear to occur in lymphocytes undergoing collective chemotaxis \cite{malet2015collective}.  Rotating droplets are, however, observed in bacteria and in the social amoeba Dictyostelium discoideum; vortex formation in bacteria has been speculated to also occur in part via chemotaxis to a secreted molecule \cite{czirok1996formation} though chemotaxis is not necessary and may not be relevant in Dictyostelium \cite{rappel1999self}.  

\subsection*{Potential extensions and comparisons to other models} 

Other variants of stochastic particle models have been used to model collective cell migration, ranging from models that use single particles to represent cells \cite{sepulveda2013collective,camley2014velocity,szabo2006phase,van2014collective,czirok1996formation} to those that use more detailed representations of cells with either multiple particles or additional details of cell shape \cite{li2014coherent,basan2013alignment,zimmermann2014intercellular,coburn2013tactile}.  Other techniques, such as the Cellular Potts Model \cite{graner1992simulation,rappel1999self,szabo2010collective} and phase field models \cite{camley2014polarity,lober2015collisions,palmieri2015multiple} have also been developed to study collective cell migration with significantly greater levels of detail on the cell's shape and its internal biochemistry.  Because emergent collective guidance has had only limited quantitative models in the past \cite{camley2015emergent,malet2015collective}, we have chosen our cell models to be as minimal as possible, in an attempt to focus on the essential aspects of collective guidance.  Earlier models have been created to study neural crest chemotaxis in vivo; however, these have explicitly assumed that chemotaxis arises from single cells following a gradient, rather than through the collective mechanism we study here \cite{mclennan2015neural}.  

We also mention that unlike many of the models discussed above, our model does not include an interaction designed to align a cell's polarity with its neighbors' motion \cite{czirok1996formation,rappel1999self,sepulveda2013collective} or its own velocity or displacement \cite{szabo2006phase,basan2013alignment,szabo2010collective,li2014coherent}, and these mechanisms are not necessary for the effects we describe here.  Competition between the collective guidance mechanism and alignment mechanisms may be an interesting area for future study.  

Our stochastic interacting particle model is relatively simple, which allows us to in some cases derive analytic results \cite{camley2015emergent}.  Many extensions of this approach are possible.  Our model could be developed further for more quantitative comparisons by careful measurement of single-cell statistics in or out of a chemoattractant gradient \cite{selmeczi2005cell,amselem2012stochastic}; this could lead to nonlinear or anisotropic terms in \eq \ref{eq:polarity}.  Our description of contact inhibition of locomotion has also assumed, for simplicity, that contact with both the front and back of the cell is inhibitory; other possibilities may alter the collective dynamics of the cell cluster \cite{camley2014polarity}.  

\subsection*{Summary}

Our main findings are: 1) Cluster velocity and chemotactic index may reflect internal signal processing, and provide an experimental window into these processes.  2) We expect sufficiently large clusters undergoing collective guidance to either become increasingly slow or break up.  3) Strong adhesion between cells is not necessary for collective guidance to function if cells chemotax to a secreted molecule. 4) A balance of this co-attraction and graded contact inhibition of locomotion are necessary for efficient chemotaxis.  5) Co-attraction may also induce cluster rotation, and we have explicitly characterized the transition to rotation.  6) The combination of cluster rotation and cluster chemotaxis may induce systematic drifts that depend on cluster rotation.

\section*{Acknowledgments}

BAC appreciates helpful discussions with Albert Bae and Monica Skoge.  This work was supported by NIH Grant No. P01 GM078586, NSF Grant No. DMS 1309542, and by the Center for Theoretical Biological Physics.  BAC was supported by NIH Grant No. F32GM110983.

\section{Appendix}
\setcounter{equation}{0}
\setcounter{figure}{0}
\setcounter{table}{0}
\renewcommand*{\thefigure}{A\arabic{figure}}
\renewcommand*{\thetable}{A\arabic{table}}
\renewcommand*{\theequation}{A\arabic{equation}}

\section{Proof of perfect adaptation and gradient sensing in limit $k_{-I}/k_D \ll 1$}

Our reaction-diffusion model (\eq \ref{eq:activator}-\eq \ref{eq:response}) for inhibitor, activator, and response on our network of cells is a direct application of the model of \cite{levchenko2002models} to a network of cells.  The steady states of \eq \ref{eq:activator} and \eq \ref{eq:response} are  $A^{i,ss} = \frac{k_A}{k_{-A}} S^i$ and
$R^{i,ss} = \frac{A^{i}/I^{i}}{A^{i}/I^{i} + k_{-R}/k_R}$.

If the signal is constant, the response steady state is independent of the signal: we can see that if $S(\rb) = S_0$, $I^{i,ss} = I^{ss}$, and that $I^{i,ss} = \frac{k_I}{k_{-I}} S_0$ and thus $A^{i,ss}/I^{i,ss} = \frac{k_A}{k_{-A}} \frac{k_{-I}}{k_I}$, independent of $S_0$.  

If the signal is {\it not} uniform, we can find the steady state of \eq \ref{eq:inhibitor} perturbatively in the limit of $\alpha \equiv k_{-I}/k_D \ll 1$.  Defining $\iota = k_I/k_{-I}$, we find that the steady state of \eq \ref{eq:inhibitor} obeys
\begin{equation}
\iota \alpha S(\ri) - \alpha I^{i,ss} + \sum_j \mathcal{L}^{ij} I^{j,ss} = 0 \label{eq:iss}
\end{equation}
where $\mathcal{L}^{ij} = C^{ij} - n^i \delta^{ij}$, where $C^{ij}$ is the adjacency matrix of the graph representing cell connections, i.e. $C^{ij} = 1$ if $i \sim j$ and $0$ otherwise.  $\mathcal{L}^{ij}$ is the ``network Laplacian" of the graph \cite{nakao2010turing}.  We note that $\sum_j \mathcal{L}^{ij} = \sum_i \mathcal{L}^{ij} = 0$.  $\mathcal{L}^{ij}$ is also a W-matrix \cite{vankampen}, and by the properties of W-matrices will have a unique (up to normalization) zero eigenvector $\sum_j \mathcal{L}^{ij} V^j = 0$, assuming that the cell cluster is connected.  This eigenvector will be constant, $V^j = 1$.  

If we write $I^{i,ss} = I^{i,ss}_{(0)} + \alpha I^{i,ss}_{(1)} + \cdots$, by equating powers of $\alpha$ we find that
\begin{align}
\sum_j \mathcal{L}^{ij} I^{j,ss}_{(0)} &= 0 \; \; \; (\textrm{Zeroth order in } \alpha) \label{eq:zeroorder} \\
\iota S(\ri) - I^{i,ss}_{(0)} + \sum_j \mathcal{L}^{ij} I^{j,ss}_{(1)} &= 0 \; \; \; (\textrm{First order in } \alpha) \label{eq:firstorder}
\end{align}
We see from \eq \ref{eq:zeroorder} and the properties of the network Laplacian discussed above that the zeroth order solution $I^{i,ss}_{(0)}$ must be a constant - $I^{i,ss}_{(0)} = \overline{I}$.  We can set the overall value of that constant by summing \eq \ref{eq:firstorder} over $i$.  $\sum_i \mathcal{L}^{ij} = 0$, leading us to conclude
\begin{equation}
\overline{I} = I^{i,ss}_{(0)} = \frac{\iota}{N} \sum_{i} S(\ri) = \iota \overline{S}
\end{equation}
where $\overline{S}$ is the mean value of $S$ over the cluster.  This result, combined with the steady-state for $A$ and the assumption that $R^{i,ss} = \frac{A^{i}/I^{i}}{A^{i}/I^{i} + k_{-R}/k_R} \approx \frac{k_R}{k_{-R}} \frac{A^i}{I^i}$ yields
\begin{equation}
R^{i,ss} \approx \frac{k_R}{k_{-R}}\frac{k_A}{k_{-A}}\frac{k_{-I}}{k_I} \frac{S(\ri)}{\overline{S}} \equiv R_0 \frac{S(\ri)}{\overline{S}}
\end{equation}
as quoted in the main paper (\eq \ref{eq:response_ss}).  

When can this be applied?  We expect that in a time $t$, $I$ will diffuse over $\sim k_D t$ cells; we expect then that if $I$ equilibrates over the cluster within the time scale $1/k_{-I}$, or $k_D /k_{-I} \gg N$, we should have good gradient sensing.  This implies that $\alpha \ll N^{-1}$ for observing linear gradient sensing.  This is merely the cluster-level version of the conditions applied for the simple one-dimensional gradient sensing module presented in in \cite{levchenko2002models}.  

\section{Details of co-attraction model}

We assume that cells secrete a chemical with concentration $c$, which diffuses in a the extracellular medium with a diffusion coefficient $D$, and breaks down with a rate $k_c$.  For a single cell at the origin, the equation for $c(\rb)$ is then:
\begin{equation}
\partial_t c(\rb,t) = D \nabla^2 c - k_c c + s\delta(\rb)
\end{equation}
where $s$ is the secretion rate.  We assume that the chemical reaches steady state, $\partial_t c = 0$.  We can solve this equation via Fourier transformation, finding that (treating our system as two-dimensional)
\begin{equation}
c(\rb) = \frac{s}{2\pi D} K_0(r/\ell)
\end{equation}
where $\ell^2 = D/k_c$ and $K_0(x)$ is the modified Bessel function of the second kind.  By superimposing many solutions, we find that for cells at positions $\rb^i$,
\begin{equation}
c(\rb^i) = \frac{s}{2\pi D} \sum_j K_0(|\rb^i-\rb^j|/\ell)
\end{equation}
Taking the gradient of this, we find (ignoring the singularity when $i = j$)
\begin{equation}
\nabla c(\rb^i) = - \frac{s}{2\pi D \ell} \sum_{j \neq i} K_1(|\vb{r}^i - \vb{r}^j|/\ell) \rij   
\end{equation}
We choose $s = 2\pi D \ell$ without loss of generality; this parameter could also be rescaled into the value of $\chi$.  

\newpage
\section{Table of parameters}

\begin{table}[h]
\begin{center}
\begin{tabular}{|c|c|c|}
\hline
Parameter symbol & Name & Value in our units \\
\hline
$\tau$ & Persistence time & 1 \\
$\sigma$ & Characteristic cell speed (OU noise parameter) & 1 \\ 
$\bar{\beta}$ & CIL strength & 20 (or as noted) \\ 
$v_a$ & Adhesion strength & 500 (or as noted) \\
$v_r$ & Cell repulsion strength & 500 (or as noted) \\
$D_0$ & Maximum interaction length & 1.2 \\
$k_A, k_{-A}$ & LEGI activator rates & Assumed fast \\
$k_R, k_{-R}$ & LEGI response rates & Assumed fast \\
$k_{I}, k_{-I}$ & LEGI inhibitor rates & 1 \\
$k_{D}$ & Cell-cell diffusion rate & 4 \\
$\xi$ & Amplification switch threshold & 0.01 \\
$S_0$ & Signal strength at origin & 1 \\
$\ell$ & Degradation length & 5 \\
$g_0$ & Gradient threshold value & $10^{-5}$ \\
$\Delta t$ & Time step & $10^{-4}$ for rigid simulations, $0.005$ for co-attraction\\
\hline
\end{tabular}
\caption{{\bf Parameters used} }
\label{tab:params}
\end{center}
\end{table}

\section*{Movie Captions}

\subsection*{S1 Video}
\label{S1_Video}
{\bf Clusters with co-attraction and graded CIL can chemotax effectively even without permanent contacts}  We show a representative movie of a chemotaxing cluster loosely bound by co-attraction, with $\betab = 70$ and $\chi = 15$.  The color map is the co-attractant field $c(\rb)$, the blue arrows are the cell polarity $\poi$, and the cells are drawn as black circles.  This video corresponds to \fig \ref{fig:coa}A.  Total time shown is 50$\tau$.

\subsection*{S2 Video}
\label{S2_Video}
{\bf Clusters with large co-attraction can develop rotational motion}  We show a representative movie of a rotating chemotaxing cluster with strong co-attraction and weaker CIL ($\betab = 37.2$ and $\chi = 83.3$).  The color map is the co-attractant field $c(\rb)$, the blue arrows are the cell polarity $\poi$, and the cells are drawn as black circles.  This video corresponds to \fig \ref{fig:coa}D.  Total time shown is 50$\tau$.




%
%
%

\end{document}